\documentclass[12pt, a4paper]{article}
\usepackage{graphicx}
\usepackage{hyperref}
\usepackage{cite}

\begin{document}
\title{Neutron form factor measurements at MAMI}
\author{B.S.~Schlimme for the A1 collaboration \\
    Institut f\"{u}r Kernphysik \\
    Johannes Gutenberg-Universit\"{a}t Mainz \\
    D-55128 Mainz, Germany}
\date{}
\maketitle

\section{Abstract}
Measurements of the electric and the magnetic neutron form factors
have been performed at the Mainz Microtron for more than 20 years.
These MAMI experiments are reviewed in the context of measurements
from other groups, and future measurements at MAMI are outlined.
%
\section{Nucleon form factors}
The electromagnetic structure of the nucleons can be probed
systematically in electron scattering experiments.  Fundamental
quantities of interest, which can be accessed in dedicated scattering
experiments, are the electric and magnetic Sachs form factors (FF)
$G_{_{\!E}}$ and $G_{_{\!M}}$ of the proton and the neutron.  The
$Q^2$ dependance of these FFs encodes the distributions of charge and
magnetization inside the nucleons. Therefore, precise measurements of
the FFs over a wide $Q^2$ range are essential for a quantitative
understanding of the nucleon structure: They are well suited for
extraction of nucleon radii and for tests of nonperturbative QCD, and
they provide constraints for phenomenological models of the nucleon
structure.  In addition, the FFs are key input to studies and searches
in particle, nuclear, and atomic physics.

From measurements of unpolarized elastic electron scattering cross
sections,
\begin{equation}
  \frac{\mathrm d \sigma}{\mathrm d \Omega}= \left(\frac{\mathrm d
    \sigma}{\mathrm d \Omega}\right)_{\mathrm{Mott}}
  \frac{1}{1+\tau}\left[G_{_{\!E}}^2(Q^2) + \frac{\tau}{\varepsilon}
    G_{_{\!M}}^2(Q^2)\right], \label{bss:eq:XS}
\end{equation}
the electric and the magnetic FFs can be extracted by an LT-separation
(i.e., measurement at a fixed $Q^2$-value, but at different
$\varepsilon$).  Here, $\left(\frac{\mathrm d \sigma}{\mathrm
  d\Omega}\right)_{\mathrm{Mott}}$ is the Mott cross section,
$\tau=\frac{Q^2}{4M^2}$, $M$ the nucleon mass, and
$\frac{1}{\varepsilon}=1+2(1+\tau)\tan^2\frac{\theta_e}{2}$ with the
electron scattering angle $\theta_e$.  Alternatively, the FFs can
directly be fitted to the measured cross sections.  In addition to
unpolarized measurements, the ratio of electric and magnetic FFs can
be accessed in double-polarization experiments.

For the proton, high precision data have been collected within many
decades (see \cite{Bernauer:2013tpr} and references therein), the
largest data set \cite{Bernauer:2010wm} comes from the Mainz Microtron
(MAMI). Despite the high quality of the available proton data and the
huge effort of the community, there are still unsolved issues.  In
particular, a significant discrepancy was found for the FF ratio
$G_{_{\!E}}^{p}/G_{_{\!M}}^{p}$ when results from unpolarized data and
from double-polarization data were compared
\cite{Punjabi:2005wq,Punjabi:2015bba}. It became clear that two photon
exchange effects are very important in order to solve that
discrepancy. However, the very details are not yet perfectly
understood (compare \cite{Henderson:2016dea}).  Also, the {\it proton
  radius puzzle} -- the discrepancy between the proton charge radius
extracted from a muonic hydrogen Lamb shift measurement and the best
present value obtained from spectroscopy of electronic hydrogen and
from elastic electron scattering experiments \cite{Bernauer:2014cwa}
-- remains unexplained and represents a serious challenge of today's
nuclear physics. Dedicated experiments are set up and performed at
various places to address that problem. Scattering experiments at low
$Q^2$ are of actual interest, because the proton charge radius can be
determined from
\begin{equation}
  r_p^2 = -6\left.\frac{\mathrm d G_{\!\mathrm E}}{\mathrm dQ^2}\right|_{Q^2=0}.
\end{equation}
At MAMI, for instance, a novel experimental technique based on initial
state radiation has been validated with the ultimate goal to precisely
measure $G_{_{\!E}}^{p}(Q^2)$ to unprecedented low $Q^2$-values
\cite{Mihovilovic:2016rkr}.

In contrast, the neutron FFs, specifically the electric neutron FF,
are much less accurately known -- due to experimental challenges.
%
\section{Challenges of neutron form factor measurements}
Measurements of the neutron FFs are dramatically hindered by the fact
that no free neutron targets with sufficient target densities are
available.  As a workaround scattering experiments on light nuclei are
performed.  For example, information on the neutron FFs can be
obtained from elastic scattering on the {\it nuclei} (see for instance
\cite{Galster:1971kv,Platchkov:1989ch} for an extraction of the
electric neutron form factor from elastic $e$-$d$ scattering data),
but large model uncertainties remain.

So to date, the best values for the neutron FFs come from {\it
  quasielastic} scattering experiments, where the electrons scatter on
the bound neutrons inside the nuclei. Compared to the {\it elastic}
scattering experiments on free protons, several severe complications
arise:

When an inclusive measurement in quasielastic kinematics is performed,
the large contributions from quasielastic scattering on the protons
must be subtracted.  Especially for a determination of the electric FF
of the neutron, which is very small compared to the other FFs, this
drastically limits the achievable accuracy.  A modern example for an
electric FF determination from an inclusive measurement can be found
in \cite{Sulkosky:2017prr}, but the gross of recent experiments
operated a (usually dedicated) neutron detector to distinguish between
scattering on the protons and on the neutrons. Detection of fast
neutrons poses a range of challenges, and so it is evident that,
depending on the concrete experiment, the performance of the neutron
detector may significantly limit the accomplishable results.

Furthermore, in quasielastic scattering one has to deal with nuclear
binding effects. The target nucleons are not free particles at rest,
but they are off-shell, and instead of the simple kinematics of a
two-body scattering process, the correlation between electron and
nucleon four-momenta is smeared out due to Fermi motion / momentum of
the residual system. This complicates the analysis of the experiment,
and it influences the quasielastic event selection capability, e.g.,
discrimination of inelastic contributions (like pion production) is
less clean.  Moreover, the extraction of the neutron FFs involves
corrections for Final State Interaction (FSI), Meson Exchange Currents
(MEC) and other effects, which can be huge.
%
\section{Magnetic neutron form factor measurements at MAMI}
The unpolarized cross section (\ref{bss:eq:XS}) for electron$-$neutron
scattering is dominated by the contribution of $G_{_{\!M}}^{n}$ due to
the smallness of $G_{_{\!E}}^{n}$. Hence, the magnetic FF can be
deduced from absolute cross section measurements by using an estimate
for the electric FF. In quasielastic scattering, however, the
scattering on the protons must be separated from the scattering on the
neutrons.

Early experiments employed {\it inclusive} quasielastic scattering on
the deuteron without explicit detection of the knocked out
neutrons. The results of these experiments were dominated by
systematic errors (specifically related to the deuteron model, FSI and
MEC corrections) rather than by statistical ones.

To minimize the sensitivity to the nuclear structure, one can extract
$G_{_{\!M}}^{n}$ from a measurement of the ratio $R$ of neutron
knockout to proton knockout cross sections in quasi-free kinematics,
\begin{equation}
R=\frac{\mathrm d \sigma(e,e'n)/\mathrm d \Omega}{\mathrm d
  \sigma(e,e'p)/\mathrm d \Omega}. \label{bss:eq:R}
\end{equation}
This ratio is insensitive to the wave function, the $e-p$ cross
section is well known, and FSI and MEC contributions are small and
well understood in the quasielastic peak \cite{Anklin:1998ae}.
%
 \begin{figure}
  \begin{center}
    \includegraphics[angle=0, width=\columnwidth]{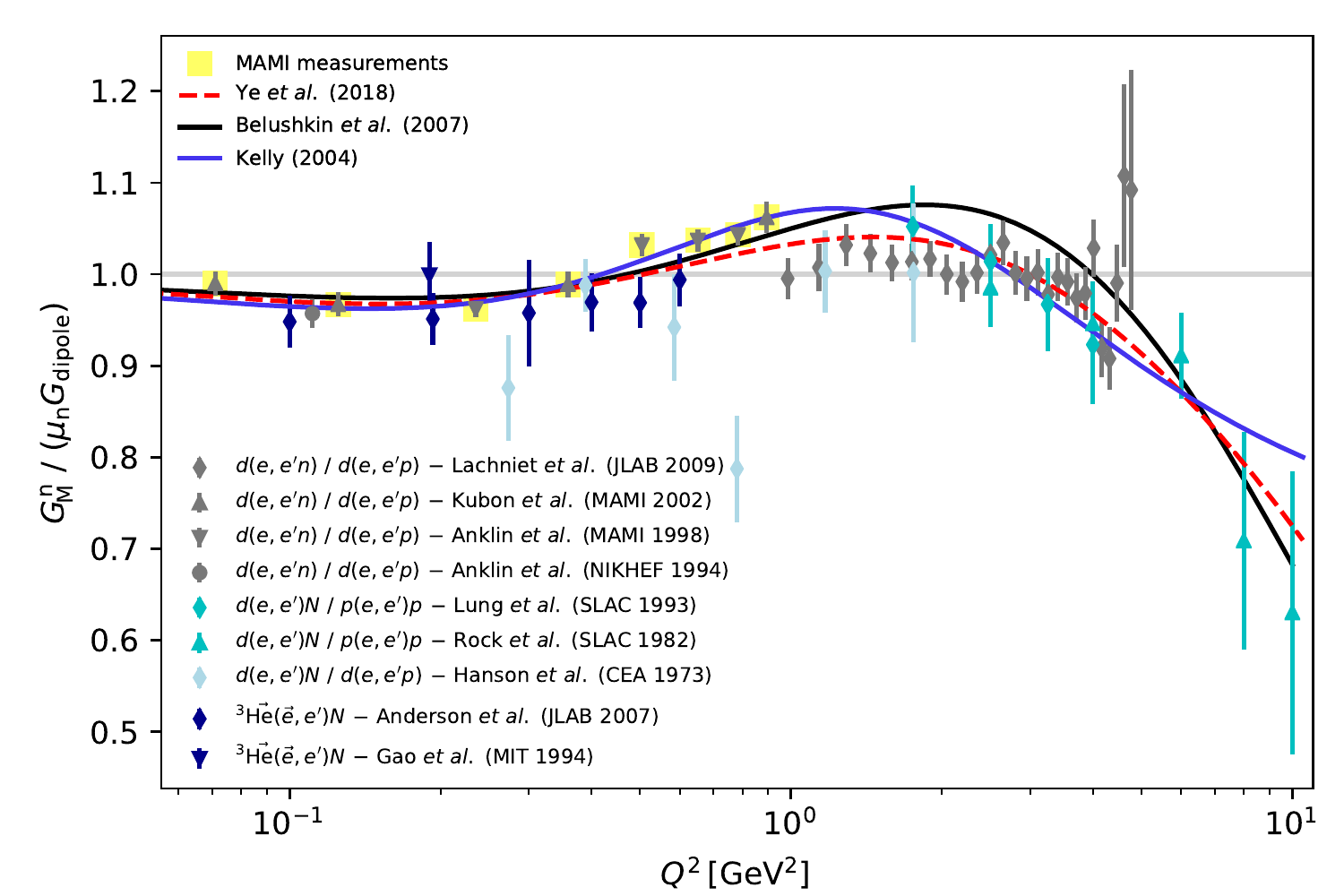}
    \caption{Selected results for $G_{_{\!M}}^{n}$ relative to the
      standard dipole FF parametrization (\ref{bss:eq:dipole})
      . Different techniques have been used: Neutron to proton
      knockout cross section ratio measurements
      \cite{Lachniet:2008qf,Kubon:2001rj, Anklin:1998ae,
        Anklin:1994ae}, inclusive measurements in the quasielastic
      regime \cite{Lung:1992bu,Rock:1982gf}, a quasi coincidence
      experiment, where the absence of a recoiling proton was
      identified as a quasielastic scattering event on the neutron
      \cite{Hanson:1973vf}, and asymmetry measurements in
      spin-dependant electron scattering on a polarized
      $^3\mathrm{He}$-target \cite{Anderson:2006jp,Gao:1994ud}.
      Measurements performed at MAMI are highlighted.  Also shown is a
      recent fit result of selected world data on $G_{_{\!M}}^{n}$ by
      \cite{Ye:2017gyb} (red line), the result of a dispersion
      analysis \cite{Belushkin:2006qa} (black line), and the
      frequently used FF parametrization of \cite{Kelly:2004hm} (blue
      line), uncertainties for these parametrizations are not
      shown.}\label{bss:fig:GMn}
  \end{center}
\end{figure}
%
Besides a few other laboratories, such measurements
\cite{Anklin:1998ae,Kubon:2001rj} were performed at MAMI
\cite{Dehn:2016mqd,Dehn:2011za,Kaiser:2008zza,Herminghaus:1976mt}. Electrons
from the MAMI accelerator impinged on a liquid deuterium target in the
spectrometer hall of the A1 collaboration \cite{Blomqvist:1998xn}. The
scattered electrons were detected with a high-resolution magnetic
spectrometer, and the ejected nucleons (neutrons as well as protons)
were detected with a nucleon detector in coincidence with the
electrons.  The simultaneous measurement of the proton and the neutron
knockout yields made the measured ratio independant of the luminosity,
dead time effects and the detection efficiency of the electron arm.
The major experimental challenge was the determination of the nucleon
detector efficiency, which is needed for the determination of the
ratio $R$ from eq. (\ref{bss:eq:R}). Especially the neutron detection
efficiency determination is difficult. It was performed using a tagged
high energy neutron beam at the Paul Scherrer Institute. Detailed
studies finally allowed to extract the magnetic form factor with an
accuracy $<2\,\%$, an order of magnitude improvement compared to the
previous determinations using inclusive measurements. With these fine
data, the authors were courageous enough to extract for the first time
a purely experimental value for the magnetic radius of the neutron
\cite{Kubon:2001rj}.

Results for $G_{_{\!M}}^{n}$ relative to the standard dipole
parametrization
\begin{equation}
  G_{\mathrm{dipole}}(Q^2) =
  \left(1+\frac{Q^2}{0.71\,\mathrm{GeV}^2}\right)^{-2} \label{bss:eq:dipole}
\end{equation}
are shown in Fig. \ref{bss:fig:GMn}.  The MAMI experiments provided
very accurate data in the important $Q^2$-range up to
$1\,\mathrm{GeV^2}$. To some extent there are tensions between the
different data sets shown in the figure, but the overall picture is
quite consistent.  In a wide $Q^2$-range, the magnetic form factor can be
well described with the dipole parametrization, some refined
parametrizations are also shown in the figure.
%
\section{Electric neutron form factor measurements at MAMI}
Measurements of $G_{_{\!E}}^{n}$ are very difficult in unpolarized
reactions since $G_{_{\!E}}^{n}$ is small due to the vanishing net
charge of the neutron, thus contributions of the neutron electric FF
to unpolarized cross sections are small.  In unpolarized quasielastic
measurements, the competing contributions from the proton and from the
neutron magnetic FF are overwhelming dominant.

From precise cross section measurements in {\it elastic}
electron$-$deuteron scattering, parametrizations for $G_{_{\!E}}^{n}$
could be obtained in a reasonable $Q^2$-range from comparison with
theoretical predictions
\cite{Galster:1971kv,Platchkov:1989ch}. Anyhow, a large model
dependance associated to the choice of the nucleon-nucleon potential
remains. In Fig. \ref{bss:fig:GEn} results from
\cite{Platchkov:1989ch} are shown for analysis using two different
$NN$ potentials (dotted gray lines).

Sensitivity to $G_{_{\!E}}^{n}$ can be tremendously enhanced in
double-polarization experiments where polarized electrons scatter
quasielastically on deuterons or $^3$He, and either the target is
polarized or the polarization of the ejected neutrons is determined
\cite{Donnelly:1985ry}.  Observables in such measurements can
particularly be sensitive to the FF ratio
$G_{_{\!E}}^{n}/G_{_{\!M}}^{n}$. Using $G_{_{\!M}}^{n}$ from other
measurements or parametrizations, $G_{_{\!E}}^{n}$ can be obtained.
These experiments are rather complex and time consuming, so the number
of existing data points is quite limited (especially when compared to
proton FF measurements), a large fraction of those comes from MAMI.
There, these measurements were started within the A3 collaboration, a
collaboration dedicated to the measurement of the electric FF of the
neutron, making use of the new technological developments which
allowed high-performance double-polarization experiments.  Two
different reactions were used, $\mathrm{\it d(\vec{\it e},{\it e'\vec
    n})p}$ and $\mathrm{^{3}\vec{He}(\vec{\it e},{\it e'n}){\it pp}}$,
but both experiments used a common detector setup. After completion of
the spectrometer hall of the A1 collaboration at MAMI
\cite{Blomqvist:1998xn}, further such experiments were performed
there.

\subsection{Neutron recoil polarization experiments}
\begin{figure}
  \begin{center}
    \includegraphics[angle=0, width=\columnwidth]{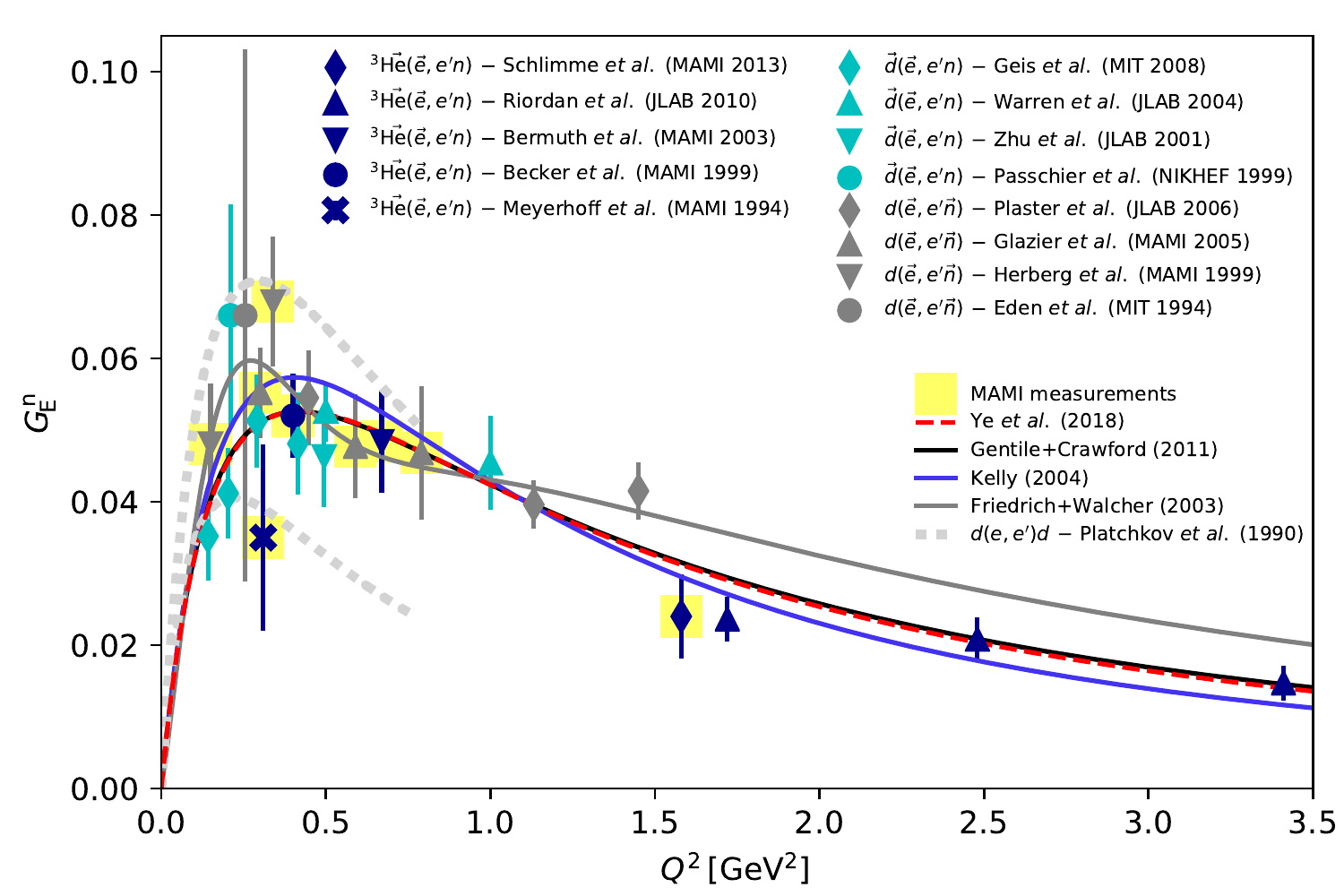}
    \caption{Results for the electric neutron FF $G_{_{\!E}}^{n}$ from
      double-polarization experiments.  All the experiments measured
      observables sensitive to the FF ratio
      $G_{_{\!E}}^{n}/G_{_{\!M}}^{n}$, $G_{_{\!E}}^{n}$ was obtained
      by inserting values for the relatively well known
      $G_{_{\!M}}^{n}$.  Different techniques have been used:
      Measurements of beam helicity asymmetries in the reaction
      $\mathrm{^{3}\vec{He}(\vec{\it e},{\it e'n}){\it pp}}$
      \cite{Schlimme:2013eoz,Riordan:2010id,Bermuth:2003qh,Becker:1999tw,Meyerhoff:1994ev},
      asymmetry measurements in $\mathrm{\vec{\it d}(\vec{\it e},{\it
          e'n}){\it p}}$
      \cite{Geis:2008aa,Warren:2003ma,Zhu:2001md,Passchier:1999cj},
      and measurements of the neutron's recoil polarization in
      $\mathrm{{\it d}(\vec{\it e},{\it e'\vec n}){\it p}}$
      \cite{Plaster:2005cx, Glazier:2004ny, Herberg:1999ud,
        Eden:1994ji}.  Measurements performed at MAMI are highlighted.
      Also shown is a recent fit result of world data by
      \cite{Ye:2017gyb} (red line), and the widely used
      parametrizations of \cite{Gentile:2011zz} (black line),
      \cite{Kelly:2004hm} (blue line) and \cite{Friedrich:2003iz}
      (gray line), uncertainties for these parametrizations are not
      shown. In addition, $G_{_{\!E}}^{n}$ results from analysis of
      precise elastic electron$-$deuteron scattering data are shown
      \cite{Platchkov:1989ch} (dotted gray lines; for two different
      nucleon-nucleon potentials used in the FF evaluation).
    }\label{bss:fig:GEn}
  \end{center}
\end{figure}

\begin{figure}
  \begin{center}
    \includegraphics[angle=0, width=\columnwidth]{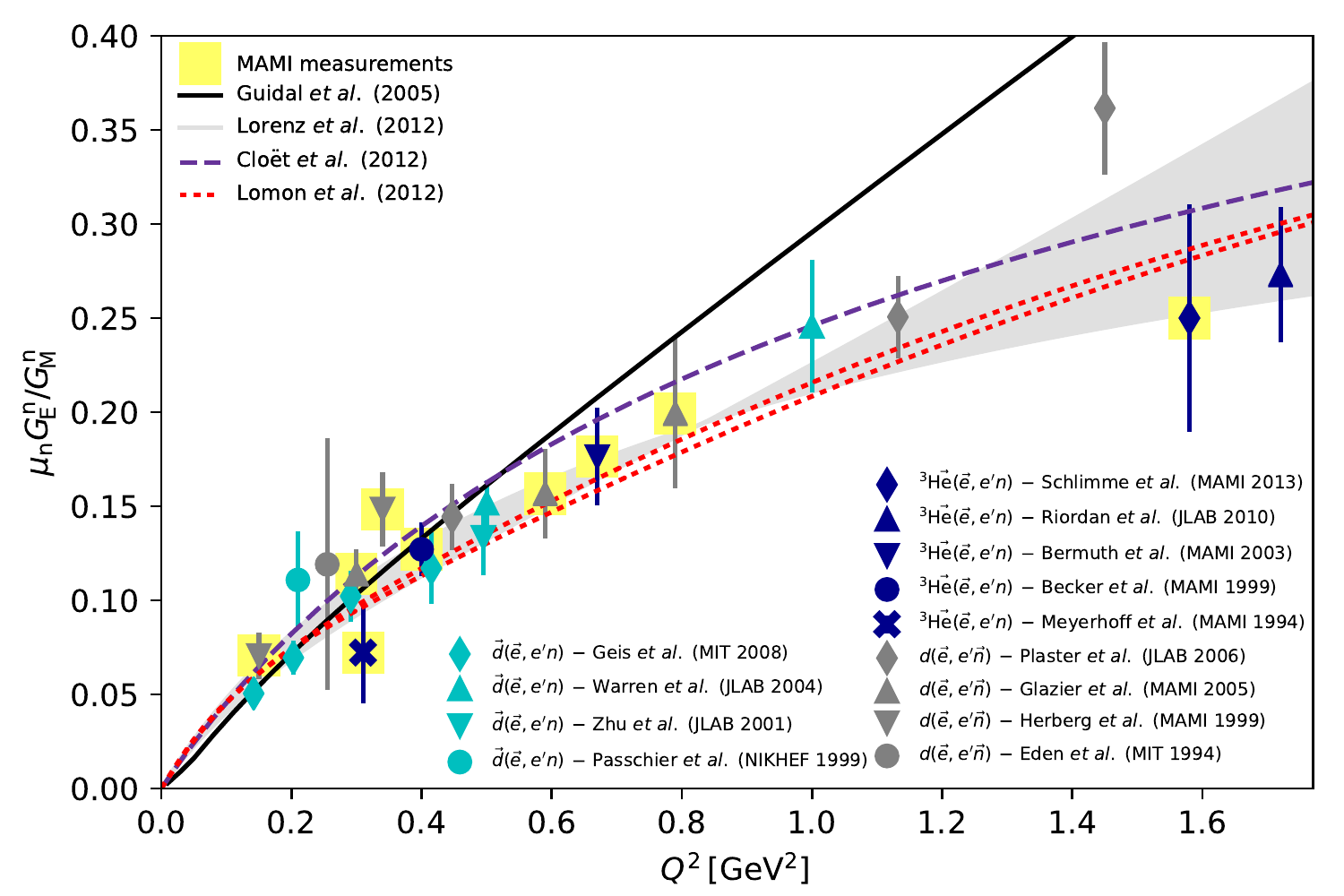}
    \caption{Results for $\mu_{\it n} G_{_{\!E}}^{n}/G_{_{\!M}}^{n}$
      \ from double-polarization experiments.  Different techniques
      have been used: Measurements of beam helicity asymmetries in the
      reaction $\mathrm{^{3}\vec{He}(\vec{\it e},{\it e'n}){\it pp}}$
      \cite{Schlimme:2013eoz,Riordan:2010id,Bermuth:2003qh,Becker:1999tw,Meyerhoff:1994ev},
      asymmetry measurements in $\mathrm{\vec{\it d}(\vec{\it e},{\it
          e'n}){\it p}}$
      \cite{Geis:2008aa,Warren:2003ma,Zhu:2001md,Passchier:1999cj},
      and measurements of the neutron's recoil polarization in
      $\mathrm{{\it d}(\vec{\it e},{\it e'\vec n}){\it p}}$
      \cite{Plaster:2005cx, Glazier:2004ny, Herberg:1999ud,
        Eden:1994ji}.  Measurements performed at MAMI are highlighted.
      Also shown are the results of recent calculations based on
      general parton distributions \cite{Guidal:2004nd} (solid line),
      dispersion analysis \cite{Lorenz:2012tm} (gray band), a
      quark-diquark model with a pion cloud \cite{Cloet:2012cy}
      (dashed line) and the extended Lomon-Gari-Kr\"umpelmann model of
      nucleon electromagnetic FF \cite{Lomon:2012pn} (dotted lines,
      for two different parametrizations of resonance widths).
      \label{bss:fig:GEnGMn}}
  \end{center}
\end{figure}

The MAMI results published in \cite{Herberg:1999ud,Glazier:2004ny}
were obtained in the reaction $\mathrm{\it d(\vec{\it e},{\it e'\vec
    n})p}$.  Polarized electrons from the MAMI accelerator impinged on
a liquid deuterium target. The helicity of the beam was reversed
during the experiments with a frequency of $1\,$Hz.  The scattered
electrons were detected in an electron detector (a leadglass-detector
array in \cite{Herberg:1999ud}, a high-resolution magnetic
spectrometer \cite{Blomqvist:1998xn} in \cite{Glazier:2004ny}).  The
neutrons were detected in coincidence with the electrons by neutron
polarimeters, charged particles were identified by thin veto
detectors.  The polarimeters provided an analysis of recoil
polarization components of the neutrons (compare
Fig. \ref{bss:fig:recoilSetup}) from which the FF ratio can be
obtained: The recoil polarization for scattering on a free neutron is
given by
\begin{eqnarray}
  P^n_x&=&
  -hP_e\cdot \frac{\sqrt{2\tau\varepsilon(1-\varepsilon)}}{\varepsilon {G_{_{\!E}}^{n}}^2 + \tau {G_{_{\!M}}^{n}}^2}\cdot  G_{_{\!E}}^{n}  G_{_{\!M}}^{n},\\
  P^n_y&=&0, \label{bss:eq:Py}\\ 
  P^n_z&=&hP_e\cdot \frac{\tau\sqrt{1-\varepsilon^2}}{\varepsilon {G_{_{\!E}}^{n}}^2 + \tau {G_{_{\!M}}^{n}}^2}\cdot {G_{_{\!M}}^{n}}^2,
\end{eqnarray}
where $h=\pm 1$ denotes the electron helicity and $P_e$ the beam
polarization degree; $\tau, \varepsilon$ as in eq. (\ref{bss:eq:XS}).
The quantization axis for $\vec P^n$ is the direction of the momentum
transfer $\vec q$.  From the ratio $P^n_x/P^n_z$ the FF ratio can be
obtained:
\begin{equation}
  \frac{P^n_x}{P^n_z} = \frac{-\sqrt{2\varepsilon}}{\sqrt{\tau(1+\varepsilon)}} \cdot \frac{G_{_{\!E}}^{n}}{G_{_{\!M}}^{n}}. \label{bss:eq:RecoilFraction}
\end{equation}
Again, nuclear binding effects have to be considered.  The effects of
FSI, MEC and isobar configuration currents (IC) on the polarization
components had been calculated based on a model
\cite{Arenhovel:1988qh}.
Corrections were found to be in the few percent range, except for the
lowest $Q^2$ measurement at $Q^2=0.15\,\mathrm{GeV^2}$, where the correction factor exceeded
$60\,\%$.  Results for these experiments are depicted in Fig. \ref{bss:fig:GEn} and Fig. \ref{bss:fig:GEnGMn}.
%
%
%
\subsection{Polarized target experiments}
FF ratio measurements can also be performed in double-polarization
experiments using polarized targets.  Besides polarized deuteron
targets, polarized $^3$He can be used as an effective polarized
neutron target due to its special spin structure, with a high relative
neutron polarization, while the mean proton polarization is small
\cite{Blankleider:1983kb,Scopetta:2006ww,Mihovilovic:2014gdi,Gentile:2016uud}.
Pioneering double-polarization measurements
\cite{JonesWoodward:1991ih,Thompson:1992ci}, performed in the {\it
  inclusive} reaction $\vec{^3{\mathrm He}}(\vec e, e')$, did not
provide very useful information on $G_{_{\!E}}^{n}$. Detection of the
recoiling neutrons could significantly improve the sensitivity of the
experiments and so, several $G_{_{\!E}}^{n}/G_{_{\!M}}^{n}$
measurements have been performed at MAMI using a polarized $^3$He
target in the reaction $\mathrm{^{3}\vec{He}(\vec{\it e},{\it
    e'n}){\it pp}}$
\cite{Schlimme:2013eoz,Bermuth:2003qh,Becker:1999tw,Meyerhoff:1994ev}.

Longitudinally polarized electrons from the MAMI accelerator were
scattered on a polarized $^3$He target (see \cite{Krimmer:2011zz} for
details on the most recent target setup).  The helicity of the beam
was reversed during the experiments with a frequency of $1\,$Hz.  The
scattered electrons were detected either in a leadglass-detector array
or, in the more recent experiments, using a high-resolution magnetic
spectrometer of the A1 collaboration \cite{Blomqvist:1998xn}, the
ejected neutrons were registered with a dedicated neutron detector
composed of scintillator bars. Beam helicity asymmetries $A$ have been
measured for different orientations of the target polarization.

\begin{figure}
  \begin{center}
    \includegraphics[angle=0, width=0.75\columnwidth]{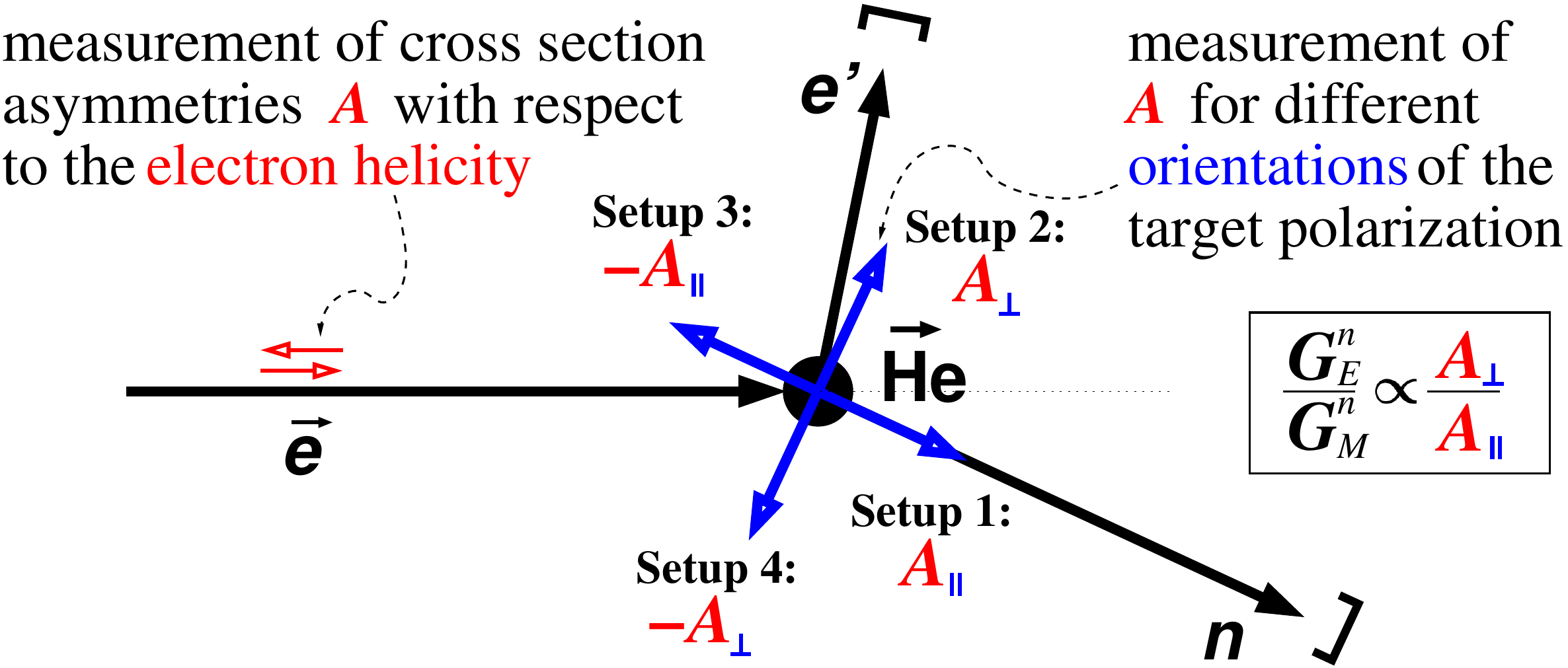}
    \caption{Asymmetries of the reaction
      $\mathrm{^{3}\vec{He}(\vec{\it e},{\it e'n}){\it pp}}$ can be
      measured for different target polarization orientations. The
      asymmetries obtained in Setup 2 and Setup 4, that correspond to
      a polarization direction in the scattering plane and
      perpendicular to the (mean) momentum transfer, are sensitive to
      $G_{_{\!E}}^{n}/G_{_{\!M}}^{n}$. Asymmetries measured in Setup 1
      and Setup 3 can be used for normalization and reduction of
      systematic errors.\label{bss:fig:MT}}
  \end{center}
\end{figure}

In the one-photon exchange approximation and for scattering on a free
neutron, $A$ is sensitive to the FF ratio for a target polarization
orientation in the scattering plane and perpendicular to the momentum
transfer $\vec q$ (see Fig. \ref{bss:fig:MT}),
\begin{equation}
  A_\perp = -\frac{a({G_{_{\!E}}^{n}/G_{_{\!M}}^{n}})}{({G_{_{\!E}}^{n}/G_{_{\!M}}^{n}})^2+d} \cdot P_{\it e}P_{\it n}.
\end{equation}
$a$ and $d$ are kinematic factors, $P_{\it e}$ and $P_{\it n}$ the
electron and neutron polarizations, respectively.  Precise measurement
of these asymmetries yield the FF ratio.
%
%
In order to minimize systematic errors, also asymmetry measurements
for a parallel target polarization orientation were performed during
all MAMI experiments. In this case, the asymmetry is almost
FF-independant,
\begin{equation}
  A_\parallel = -\frac{b}{({G_{_{\!E}}^{n}/G_{_{\!M}}^{n}})^2+d} \cdot P_{\it e}P_{\it n},
\end{equation}
because of the smallness of the ratio,
$({G_{_{\!E}}^{n}/G_{_{\!M}}^{n}})^2\ll d$. $b$ is another kinematic
factor.  The ratio of these asymmetries leads to
\begin{equation}
  G_{_{\!E}}^{n}/G_{_{\!M}}^{n}=b/a \cdot \frac{\left(P_eP_n\right)_\parallel}{\left(P_eP_n\right)_\perp}A_\perp/A_\parallel \label{bss:eq:g}
\end{equation}
with great benefits: calibration factors for the beam and target
polarization measurements cancel, and only a relative monitoring is
necessary. In particular, the precise value for the relative neutron
to helium polarization (smaller than 1, and only the helium
polarization can be measured directly) is irrelevant. Also,
unpolarized background drops in the ratio. That is of special interest
concerning unavoidable background from quasielastic electron$-$proton
scattering (this is, for instance, the dominant contribution to the
systematic error in \cite{Glazier:2004ny}): The polarization of the
protons is small inside the $^3$He nucleus, and so their contribution
is almost unpolarized and drops out in the asymmetry ratio.

Results for these experiments are also shown in Fig.
\ref{bss:fig:GEn} and Fig. \ref{bss:fig:GEnGMn}.  Since
eq. (\ref{bss:eq:g}) is only valid for scattering on free neutrons,
corrections for nuclear effects had to be applied. The earliest of
these experiments \cite{Meyerhoff:1994ev} was uncorrected for nuclear
physics effects, which possibly explains the deviation from other
data; the result is included in the figures only for
completeness. This experiment was later repeated with much better
statistics, the ``raw'' result from this experiment
\cite{Becker:1999tw} was corrected for FSI and MEC contributions in an
independant publication \cite{Golak:2000nt} (where the systematic
experimental error was ignored), based on Faddeev calculations.  The
data point shown in the figures corresponds to the FSI-corrected
result from \cite{Golak:2000nt} but with the relative systematic error
from the original paper \cite{Becker:1999tw} added for a fair
representation of the result of that experiment.  For the analysis of
\cite{Bermuth:2003qh} corrections for FSI were estimated by scaling
available calculations of \cite{Golak:2001ge}, performed at a smaller
$Q^2$, to the $Q^2$ of the experiment. This resulted in a total FSI
correction of $3.4\,\%$.  For the kinematics of the experiment at
$Q^2=1.58\,\mathrm{GeV^2}$ \cite{Schlimme:2013eoz}, no significant FSI
and MEC effects were expected, calculations based on the generalized
eikonal approximation \cite{Sargsian:2004tz} were performed and
confirmed that.  As a fact, further measurements had been performed at
MAMI using a polarized $^3$He target at lower momentum transfers of
$Q^2=0.25\,\mathrm{GeV^2}$ (see \cite{Grabmayr:2008zz}) and even at
$Q^2=0.15\,\mathrm{GeV^2}$. The necessary corrections due to FSI and
MEC were found to be huge and non-constant over the detector
acceptances causing severe problems (especially for the lowest $Q^2$
measurement), the analyses have not been finalized. \\

Taking all the dedicated double-polarization experiments together, the
shape of $G_{_{\!E}}^{n}(Q^2)$ has been nicely elaborated, with strong
contributions coming from MAMI. With the experiments mentioned above,
the full $Q^2$-range accessible at MAMI has been exploited: At
smallest $Q^2$ ($\leq 0.2\,\mathrm{GeV^2}$), analyses become
unreliable as a result of the increasing size of nuclear binding
effects on the observables. A significant increase of the momentum
transfer beyond $1.6\,\mathrm{GeV^2}$ is not possible due to the
limited beam energy of the accelerator ($1.6\,\mathrm{GeV}$).
%
\section{Planned measurements}
In the light of the initial difficulties to get a handle on the
neutron electric FF, it is in a way astonishing to see how well all
these $G_{_{\!E}}^{n}$ measurements from different laboratories, from
the study of different reactions, using different approaches for
nuclear binding effects corrections, align. Considering that the
necessary corrections to the data are partly excessive and that
eventually not all relevant effects have been accounted for (maybe the
structure of a free neutron, a neutron bound in a deuteron, and a
neutron bound in a $^3$He-nucleus is not identical but differs
significantly? see \cite{Yaron:2016dal,Izraeli:2017kgg} for
investigations of that subject in the proton case), the diversity of
the experiments is excellent. Clearly, the data quality is different
than in the case of the proton, but alltogether the global efforts of
experimentalists, engineers, and theoreticians have been worth it.

However, there are also limitations with the existing data sets.  How
to interprete deviations of the data, for instance in the region
$Q^2\approx 1.5\,\mathrm{GeV^2}$? Revealing structures in the FF is
difficult when data from different experiments, each single one
featuring its own systematics, are compared.

\begin{figure}
  \begin{center}
    \includegraphics[angle=0, width=\columnwidth]{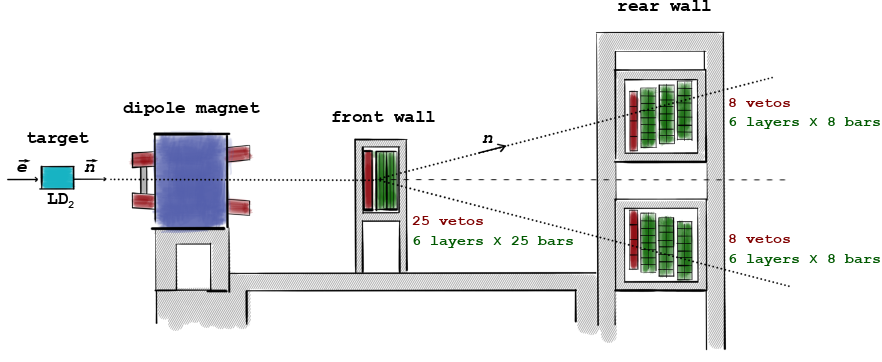}
    \caption{Sketch of a neutron polarimeter for neutron FF
      measurements at MAMI (side view).  Spin dependant scattering
      takes place in the detector material of the front wall, leading
      to an asymmetry in the azimuthal angle of the secondary
      scattered neutrons.  The analysis scattering is only sensitive
      to the transverse neutron spin components.  A non-vanishing
      neutron spin component perpendicular to the neutron momentum
      direction in the horizontal plane leads to an ``up-down''
      asymmetry. This asymmetry can be measured by use of the rear
      wall.  Frequent reverses of the electron beam helicity provide
      suppression of false asymmetries related to a non-constant
      detector efficiency and so forth.  A vertical spin component
      would cause a ``left-right'' asymmetry, but as result of
      eq. (\ref{bss:eq:Py}) that component approximately vanishes in
      electron$-$neutron scattering.  By means of a dipole in front of
      the detectors, the neutron spins can be rotated about a vertical
      axis, making the polarimeter setup sensitive to the longitudinal
      spin component of the incident neutrons. The analyzing power of
      the detector material does not change, and therefore largely
      cancels in the ratio of the measured polarization components,
      together with the absolute value of the electron beam
      polarization.  }\label{bss:fig:recoilSetup}
  \end{center}
\end{figure}

In that sense a complementary measurement program is planned at MAMI
with the aim to collect a consistent data set over a wide $Q^2$-range
($Q^2=0.2 - 1.5\,\mathrm{GeV^2}$) using the $\mathrm{\it d(\vec{\it
    e},{\it e'\vec n})p}$ reaction as described before. For this
purpose, a new highly segmented neutron polarimeter is being set up,
see Fig. \ref{bss:fig:recoilSetup}.  The neutron polarimeter will be
operated in conjunction with the high-resolution magnetic
spectrometers of the A1 collaboration for precise electron detection.

The recoiling neutrons will pass a dipole magnet, where the vertical
field can be used to precess the neutron spins about a vertical
axis. Analysis scattering of the neutrons takes place in the front
wall of the neutron detector in vertically aligned plastic
scintillator bars (scintillator: EJ-200, PMT: 9142SB). For charged
particle identification, thin veto layers are placed in front of the
wall. 

The scintillators of a second wall will be arranged horizontally in
two blocks above and below the electron scattering plane. Study of
$\sim$up-down asymmetries for different settings of the dipole magnet
allows the extraction of the recoil polarization components $P^n_x$
and $P^n_z$ of the neutrons, the fraction of those components yields
$\frac{G_{_{\!E}}^{n}}{G_{_{\!M}}^{n}}$, compare
eq. (\ref{bss:eq:RecoilFraction}).
%
%
\section{Summary}
Nucleon electromagnetic FFs are fundamental quantities
describing the electromagnetic structure of the proton and the
neutron.  These FFs can be determined in electron scattering
experiments. A serious complication for the measurements of the
neutron FFs is the lack of free neutron targets, the FFs are
extracted from measurements on light nuclei therefore. Contributions
related to the protons inside the nuclei limit the accuracy of the
experiments, especially the electric neutron form factor is hard to 
measure because it is small compared to the other form factors, and so its
contribution to the observables is in general small, too.

With technological improvements, sophisticated experiments could be
set up. Continuous-wave accelerators provided excellent electron
beams, high-resolution magnetic spectrometers could be used for
precise electron detection, the recoiling neutrons could be detected
in coincidence, allowing a clean selection of quasielastic
electron$-$neutron scattering events. In addition, polarization
degrees of freedom became available and were proofed to be very
beneficial in order to access the electric neutron FF.

A substantial number of data points has been obtained at the Mainz
Microtron.  Both the magnetic and the electric FF were measured at
various momentum transfers in the range $Q^2=0.15-1.58\,\mathrm{GeV^2}$.
The magnetic FF was determined from measurements of the neutron to
proton cross section ratio in quasielastic electron scattering, the
dominant error source -- the neutron detector efficiency -- was
encountered by supplemental efficiency measurements using a tagged
neutron beam.  The electric FF was determined from measurements of the
electric to magnetic FF ratio combined with measurements of the
magnetic FF. Utilizing a polarized electron beam, this ratio was
either determined from a measurement of the neutron recoil
polarization components, or from the study of beam helicity
asymmetries in scattering on an effective polarized neutron target.

Although the quality of the available neutron FF data, including the
significant contributions from MAMI, is of course not as good as it is
for the proton FFs, the situation is quite satisfactory.  The shape of
both magnetic and electric FF has been worked out well with some space
for interpretation.

New recoil polarization measurements are planned, with the goal to
consistently scan the electric FF over the $Q^2$-range
accessible at MAMI.
%
\section*{Acknowledgements}
We gratefully acknowledge the support from the technical staff at the
Mainz Microtron and thank the accelerator group for the excellent beam
quality for more than two decades.  This work was supported by the
Deutsche Forschungsgemeinschaft with the Collaborative Research Center
1044.
%

\end{document}